\documentclass[twocolumn,showpacs,aps,amsmath,amssymb]{revtex4}

\usepackage{epsfig}

\def\be{\begin{equation}}
\def\ee{\end{equation}}
\def\bea{\begin{eqnarray}}
\def\eea{\end{eqnarray}}
\def\bes{\begin{subequations}}
\def\ees{\end{subequations}}

\def\beq{\begin{equation}}
\def\eeq{\end{equation}}
\def\barr{\begin{eqnarray}}
\def\earr{\end{eqnarray}}

\def\bb{\alpha_2}
\def\cc{\alpha_3}

\def\lsim{\:\raisebox{-1.1ex}{$\stackrel{\textstyle<}{\sim}$}\:}
\def\gsim{\:\raisebox{-1.1ex}{$\stackrel{\textstyle>}{\sim}$}\:}

\begin{document}

\title[]{Quark-lepton complementarity with quasidegenerate
Majorana neutrinos} 
 
\author{Amol Dighe$^1$, 
Srubabati Goswami$^{2,3}$ ~and
Probir Roy$^{1}$
}
\affiliation{
1. Tata Institute of Fundamental Research, 
Homi Bhabha Road, Mumbai 400 005, India \\
2. Harish-Chandra Research Institute, Chhatnag Road, Jhusi,
Allahabad 211 019, India \\
3. Technische Universit\"at
M\"unchen, James--Franck--Strasse, D--85748 Garching, Germany
}

\begin{abstract}
A basis independent formulation of quark-lepton complementarity is
implemented at a high scale for quasidegenerate 
Majorana neutrinos.
It is shown that even with the renormalization group evolution
in the minimal supersymmetric standard model,
the scenario can be consistent with the data provided a nontrivial
role is played by the Majorana phases.
Correlated constraints are found on these phases and the neutrino 
mass scale using the current data.
We also indicate how future accurate measurements of the 
mixing angles can serve as tests of this scenario and restrict the values
of the Majorana phases.
\end{abstract}


\maketitle

Neutrinos provide a fertile ground for novel and testable ideas 
due to the present availability and future prospects of more
precise information \cite{nu-fits} 
on their masses and mixing angles. We
aim to combine three such ideas in this letter: (a) quark-lepton
complementarity (QLC) 
\cite{raidal,minakata,giunti,qlc,antusch-moha,smirnov05}, 
(b) quasidegenerate neutrinos (QDN) \cite{qdn},
and (c) nontrivial Majorana phases.
The possibility of such a combination was mentioned in \cite{minakata},
but we explicitly demonstrate its feasibility 
by use of the renormalization group (RG) evolution
in a transparently analytic way.

QLC links the difference between the maximal ($45^\circ$) and
the measured ($33.8^{+2.4}_{-1.8}$ degrees \cite{nu-fits})
value of the neutrino mixing angle $\theta_{12}$
to the Cabibbo angle 
$\theta_c = 12.6^\circ \pm 0.1^\circ$ \cite{pdg}. 
This can be done by postulating the relation
$\theta_{12} + \theta_c = 45^\circ$.
However, the various implementations of this relation
in the literature (e.g. in \cite{minakata}) are
fraught with basis ambiguities \cite{jarlskog} 
and issues of scale \cite{antusch-moha}.  
In this letter, we follow a basis independent formulation 
of QLC from \cite{minakata}:
\be
U_{PMNS} = V^\dagger_{CKM} U^{\rm bimax}_\nu \; ,
\label{vdagv}
\ee
where $U_{PMNS}$ is the Pontecorvo-Maki-Nakagawa-Sakata matrix
unitarily transforming mass eigenstates of neutrinos to their flavor
eigenstates, $V_{CKM}$ is the Cabibbo-Kobayashi-Maskawa
matrix \cite{pdg} and $U^{\rm bimax}_\nu$ is the unitary matrix which
diagonalizes the bimaximal form \cite{bimaximal} of the neutrino Majorana mass
matrix ${\cal M}^{\rm bimax}_\nu$.  
Eq.~(\ref{vdagv}) yields 
\beq
\theta_{12} + \theta_c/\sqrt{2} = 45^\circ + {\cal O}(\theta_c^2) \; .
\label{qlc-form}
\eeq

The identification of eq.~(\ref{vdagv}) as a statement of QLC
becomes more transparent in the basis with $U_u = 1$, where
$U_f$ represents the unitary (mass $\rightarrow$ flavor)
transformation of the left chiral components of the $f$ $(= u,d,l)$
type of charged fermions which diagonalize the Yukawa 
coupling matrix combination $Y_f^\dagger Y_f$ \cite{chankowski}.
Thus, in this basis, $Y_u^\dagger Y_u$ is diagonal in flavor
space.  It follows that $V_{CKM}$ $(\equiv U^\dagger_u U_d)$ now
equals $U_d$, so that a comparison between eq.~(\ref{vdagv}) and the
definition of $U_{PMNS}$ $(\equiv U^\dagger_l U_\nu)$, with the
assumption of $U_\nu$ being $U^{\rm bimax}_\nu$, now yields the
quark-lepton symmetry relation $U_d = U_l$.  Eq.~(\ref{vdagv}), as it
stands, is basis independent, however.

A quark-lepton symmetry relation such as (\ref{qlc-form}) 
is expected to be valid at the GUT scale $\sim 10^{16}$ GeV.
In our scenario, neutrino masses are generated by an
effective dimension-5 operator ${(l \cdot h)(l \cdot h)/\Lambda}$
at the scale $\Lambda$ and the mechanism
that gives rise to this operator is immaterial.
The mechanism may include right handed neutrinos, in which case
we have to assume that the threshold effects \cite{parida}
do not spoil the relation till $\Lambda \sim 10^{12}$ GeV,
above which all the right handed neutrinos are expected to lie.
All the other threshold effects are taken to be flavor blind.
We thus postulate the relation (\ref{qlc-form}) to hold at a scale 
$\Lambda \sim 10^{12}$ GeV.
Our results are only logarithmically sensitive to the
exact choice of scale.

Quasidegenerate neutrinos are very much allowed by present 
cosmological constraints \cite{cosmology} as well as neutrinoless double 
$\beta$-decay experiments \cite{double-beta}.
From the model building point of view, the quasidegenerate spectrum 
can be obtained rather naturally through type II seesaw
mechanism by invoking discrete symmetries like
flavor $SO(3)$ \cite{qdn}.
It can also be produced in models with Abelian family symmetries 
\cite{binetruy}, or with flavor symmetries like 
$L_\mu-L_\tau$ \cite{choubey}.
In the minimal supersymmetric standard model (MSSM), the
neutrino masses and mixing angles 
may evolve significantly from $\Lambda$ to the SUSY breaking scale 
$\lambda$ via the RG equations \cite{rg-papers,antusch-majorana}. 
This evolution can potentially spoil the QLC signatures in the low
energy data \cite{minakata}.
In this letter, we study the evolution of the QLC equation
(\ref{vdagv}) analytically as well as numerically,
including the effect of Majorana phases
\cite{antusch-majorana,haba}, and show its consistency with the 
observed mixing angles in the QDN scenario.

We take the neutrino masses $m_{1,2,3}$ to be complex in general
and parametrize their absolute values
in terms of three real parameters $m_0,\rho_A$ and
$\epsilon_S$ as 
\bea 
|m_1| &=& m_0 (1 - \rho_A) (1 - \epsilon_S) \; , \nonumber \\
 |m_2| &=&m_0 (1 - \rho_A) (1 + \epsilon_S) \; , \nonumber \\
 |m_3| &=& m_0 (1 + \rho_A),
\label{rhoeps-def}
\eea 
with $m_0$, the parameter setting the neutrino mass scale,
and $\epsilon_S$ being required to be positive,
while a positive
(negative) $\rho_A$ implies a normal (inverted) neutrino
mass ordering.  
These parameters may be related to the solar and atmospheric
mass squared differences 
($\delta m^2_S \sim 8 \times 10^{-5}$ eV$^2$ and $|\delta m^2_A| \sim
2 \times 10^{-3}$ eV$^2$)
and the the sum of the neutrino absolute masses through
\bea
\label{dmsq-rhoeps}
\delta m^2_S &=& |m_2|^2 - |m_1|^2 \approx 4m^2_0 (1 - \rho_A)^2 
\epsilon_S \; ,
\phantom{+ {\cal O}(\epsilon^2_S)} \nonumber \\
|\delta m^2_A| &=& ||m_3|^2 - (|m_1|/2 + |m_2|/2)^2 | = 
4 m^2_0 |\rho_A| \; , \nonumber \\
\Sigma_i |m_i| & = & 3m_0 (1 - \rho_A/3) \; .
\eea
In (\ref{dmsq-rhoeps}) we have made use of $|\epsilon_S| \ll 1$
(for instance, if $m_0 \simeq 0.2$ eV, one has 
$|\rho_A| \simeq 1.8 \times 10^{-2}$ and 
$\epsilon_S \simeq 5 \times 10^{-4}$) 
while neglecting terms which are ${\cal O}(\epsilon_S^2)$.

The RG evolution of the hierarchical charged fermion masses
is known to be small \cite{rg-charged}, and we neglect it.
The bimaximal neutrino mass matrix emerging at a high scale $\Lambda$ 
gets modified at the low scale $\lambda$ to yield 
\cite{chankowski,ellis-lola}
\beq
M_\lambda = {\cal I}_K ~ {\cal I}_\kappa^T ~ {\cal M}_\nu^{\rm bimax} 
~{\cal I}_\kappa \; ,
\label{m-lambda}
\eeq
where 
${\cal I}_K \equiv \exp[ - \int_{t(\Lambda)}^{t(\lambda)} K(t) dt]$
is the scalar factor that arises from the RG evolution due to the
gauge couplings and the fermion-antifermion loops.
In MSSM, we have
$K(t) = -6 g_2^2 -2 g_Y^2 + 6 {\rm Tr}(Y_u^\dagger Y_u)$.
Here $t(Q)$ is defined to be
$t(Q) \equiv (16 \pi^2)^{-1}  \ln(Q/Q_0)$
with $Q_0$ an arbitrary scale.

The other factor ${\cal I}_\kappa$ in (\ref{m-lambda}) is given by
\beq
{\cal I}_\kappa \equiv \exp [ -  
\int_{t(\Lambda)}^{t(\lambda)} (Y_l^\dagger Y_l)(t) dt ] \; .
\eeq
In the basis chosen for our QLC scenario, 
\beq
Y_l^\dagger Y_l = V_{CKM} 
\mbox{ Diag}(y_e^2, y_\mu^2, y_\tau^2)~
V_{CKM}^\dagger \; .
\eeq
Since $y_e^2 \ll y_\mu^2 \ll y_\tau^2$, we can neglect $y_e$ and
$y_\mu$.
If, in addition, we neglect the elements of the CKM matrix that are
of the order of $\theta_c^2$ or smaller, only the \{3--3\} element 
of $Y_l^\dagger Y_l$ survives. Then we have
\beq
{\cal I}_\kappa \approx 
\mbox{ Diag} (1,1,e^{-\Delta_\tau}) \; ,
\eeq
where in the MSSM, one has
\beq
\Delta_\tau = m_\tau^2 (\tan^2 \beta+1) (16 \pi^2 v^2)^{-1} 
\ln (\Lambda/\lambda) \; .
\label{delta-tau}
\eeq
Here 
$v \equiv \sqrt{v_u^2 + v_d^2}$ where $v_u$ 
and $v_d$ are the vevs of the two neutral Higgs scalars, with
$\tan\beta \equiv v_u/v_d$. 
For $\Lambda \sim 10^{12}$ GeV, $\lambda \sim 10^3$ GeV, 
$\tan\beta \sim 30$ and $v \sim 246$ GeV, we find that
$\Delta_\tau \sim 6 \times 10^{-3}$.
Therefore, unless the coefficients of $\Delta_\tau$ are
${\cal O}(10^2)$ or higher, we can neglect terms that 
involve two or more powers of $\Delta_\tau$.
Henceforth, we keep only the terms linear in $\Delta_\tau$.

The mass matrix (\ref{m-lambda}) in the flavor basis
takes the following form at the low scale: 
\beq
M_\lambda  =   
\left( \begin{array}{ccc}
A & B & -B X \\
B & C + A/2 & (C-A/2) X \\
-B X & (C-A/2) X & (C+A/2) Y \\
\end{array} \right) {\cal I}_K
\label{m-low}
\eeq
where we have used the notation
$A \equiv (m_1+m_2)/2 \; , 
B \equiv (-m_1 + m_2)/(2\sqrt{2}) \; ,
C \equiv m_3/2 \; ,
X \equiv (1-\Delta_\tau) \;,$ and
$Y \equiv (1- 2\Delta_\tau)$.
 
We parametrize the unitary matrix $U_\lambda$ that diagonalizes 
$M_\lambda$ by
\barr
U_\lambda & \equiv & 
{\rm Diag}(e^{i\phi_e \Delta_\tau}, e^{i\phi_\mu \Delta_\tau}, 
e^{i\phi_\tau \Delta_\tau})
R_{23}(\pi/4+k_{23}\Delta_\tau) \times \nonumber \\
& & {\rm Diag}(1, 1, e^{i \delta}) 
R_{13}(k_{13} \Delta_\tau) 
{\rm Diag}(1, 1, e^{-i \delta}) \times \nonumber \\
 & & R_{12}(\pi/4+k_{12}\Delta_\tau)
{\rm Diag}(e^{i\alpha_1/2}, e^{i\alpha_2/2}, e^{i\alpha_3/2}) \, , 
\earr
where $R_{ij}$ is the rotation matrix in the $ij$ plane,
$\alpha_i$'s are the Majorana phases, and the phases
$\phi_e$, $\phi_\mu$ and $\phi_\tau$ are required to
diagonalize a general neutrino mass matrix
\cite{Antusch:2005gp}.
Thus the new mixing angles are
\bea
\theta_{12} = \pi/4 + k_{12} \Delta_\tau \; ,
\theta_{23} = \pi/4 + k_{23} \Delta_\tau \; ,
\theta_{13} = k_{13} \Delta_\tau  \;.
\nonumber
\label{thetaijs}
\eea
For $\Delta_\tau =0$, we have $U_\lambda = U_\nu^{\rm bimax}$. 
We approximate the deviation of $U_\lambda$ from $U_\nu^{\rm bimax}$ 
by keeping terms that are linear in $\Delta_\tau$.
The current allowed ranges of the mixing angles are such that the
deviations from QLC values without RG running are very small.
Therefore, the approximation $|k_{ij} \Delta_\tau| \ll 1$ 
should be always valid so that we can neglect the higher order terms 
in $k_{ij} \Delta_\tau$.
Furthermore, the Dirac phase $\delta$, which is vanishing at the
high scale and is generated only through the RG evolution,
is retained only to the first order, and consequently
plays no role in our ${\cal O}(\Delta_\tau)$ analysis. 

The values of $k_{ij}$ are found to be
\barr
k_{12} & = & \frac{1}{4} \frac{|m_1 + m_2|^2}{(|m_2|^2 - |m_1|^2)} 
\; , \nonumber \\
k_{23} & = & \frac{1}{4} \left[ 
\frac{|m_2 + m_3|^2}{(|m_3|^2 - |m_2|^2)} +
\frac{|m_1 + m_3|^2}{(|m_3|^2 - |m_1|^2)} \right] \;, \nonumber \\
k_{13} & = & \frac{1}{4} \left[ 
\frac{|m_2 + m_3|^2}{(|m_3|^2 - |m_2|^2)} -
\frac{|m_1 + m_3|^2}{(|m_3|^2 - |m_1|^2)} \right] \; .
\phantom{space}
\label{kij-m1m2m3}
\earr
The above expressions are valid as long as the values of
$m_i$'s and $\delta m_{S/A}^2$'s are described accurately by
the ${\cal O}(\Delta_\tau)$ terms in their RG evolution.
Though this condition always holds with the $m_i$'s, for
$\Delta_\tau \gsim \delta m^2_{S/A}/m_0^2$ the ${\cal O}
(\Delta_\tau^2)$ terms dominate over the ${\cal O}
(\Delta_\tau)$ terms in $\delta m_{S/A}^2$
\cite{antusch-majorana} and eqs.~(\ref{kij-m1m2m3})
are no longer a good approximation.
They are valid only for
$\Delta_\tau \lsim \delta m_S^2/m_0^2$, i.e.
for $m_0 \tan \beta \lsim 3$ eV.
Higher $\Delta_\tau$ values also lead to 
$|k_{12} \Delta_\tau| \gg 1$,
so that the evolution of $\theta_{12}$ is too large to be 
naturally accomodated with the current data.

Equations~(\ref{kij-m1m2m3}) 
are consistent with the running of angles
computed in the general case \cite{antusch-majorana},
though they have been computed here in a much simpler way for
the special case of bimaximal neutrino mixing.
In terms of the parameters $m_0, \rho_A, \epsilon_S$ defined in
eq.~(\ref{rhoeps-def}), the expressions (\ref{kij-m1m2m3}) become
\barr
k_{12} & = & 
[(1+\cos\bb)+\epsilon_S^2(1-\cos\bb)] 
/(8 \, \epsilon_S) \; , \nonumber \\
k_{23} & = & \frac{\Gamma}{8}
[2 +  \cos(\bb-\cc) + \cos\cc ] + \frac{\rho_A}{2}+ {\cal O}(\epsilon_S) \; , 
\phantom{spa} \nonumber \\
k_{13} & = & (\Gamma/8) 
[ \cos(\bb-\cc) - \cos\cc ] + {\cal O}(\epsilon_S) \; ,
\label{kij-rhoeps}
\earr
where $\Gamma \equiv (1/\rho_A) - \rho_A$.
One of the three Majorana phases $\alpha_{1,2,3}$ can be rotated away
and we have chosen that to be $\alpha_1$.

The following observations can now be made:\\ 
\noindent 
$\bullet$ Quasi-degeneracy of the neutrinos 
($m_0^2 \gg \delta m^2_{S/A}$) is required for any significant 
enhancement of all $k_{ij}$'s: the magnitude of $k_{12}$ is
enhanced when $\epsilon_S = \delta m^2_S/[4 m_0^2 (1-\rho_A)^2] 
\ll 1$, whereas the magnitudes of
$k_{23}$ and $k_{13}$ are enhanced for 
$\rho_A \delta m^2_A/(4 m_0^2) \ll 1$.

\noindent $\bullet$
The values of the Majorana phases are crucial in deciding 
the values of $k_{ij}$'s: 
As $\bb \to 0$ we have $k_{12} \approx 1/(4 \epsilon_S)$.
When $\bb$ is nonzero, the value of $k_{12}$ decreases rapidly.
At $\bb = \pi$, we have $k_{12}$ to be nearly as small as 
$\epsilon_S/4$.
Both $|k_{23}|$ and $|k_{13}|$ are enhanced when $\cc =0$ or $\cc=\bb$. 
However, when $\bb =0$, the magnitude of $k_{13}$ is highly 
suppressed.

\noindent $\bullet$
$k_{12}$ is independent of $m_3$ as well as $\cc$, 
and is always positive. 
$k_{23}$ is positive (negative) for the normal (inverted) 
neutrino mass ordering.
The sign of $k_{13}$ depends on the ordering as well as
the Majorana phases.

\begin{figure}
\epsfig{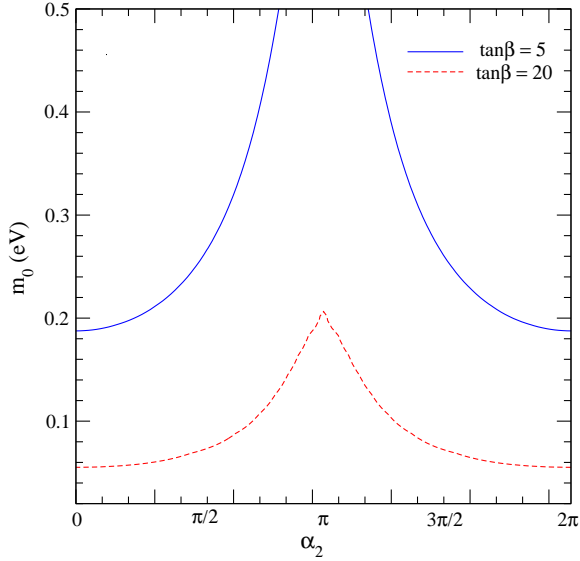}
\caption{Contours of $\theta_{12}$ =39.8$^\circ$  in the
$m_0$ (eV)--$\bb$ (radians) plane shown for $\tan\beta$ = 5 (20)
by  solid (dashed) lines.
The regions above the contours are excluded by data for that
particular $\tan\beta$ value.
\label{fig-mbeta}
}
\end{figure}

The net leptonic mixing matrix at the low scale is 
$V_{\rm PMNS} = 
V_{\rm CKM}^\dagger U_\lambda$ with the mixing angles 
given by $\theta_{ij} = \theta_{ij}^0 + k_{ij} \Delta_\tau$,
where
$\theta_{12}^0 \approx 35.4^\circ, 
\theta_{23}^0 \approx 42.5^\circ, 
\theta_{13}^0 \approx 8.9^\circ$
are their QLC values at the high scale.
Whereas $\theta_{12}^0$ and $\theta_{13}^0$ are known to an 
accuracy of $\approx \pm 0.1^\circ$,
the exact value of $\theta_{23}^0$ 
depends on the value of the $CP$ violating phase $\delta$
in the CKM matrix \cite{minakata},
and is currently uncertain by nearly $\pm 1^\circ$.
The ``deviations'' $\Delta \theta_{ij} \equiv 
\theta_{ij}-\theta_{ij}^0 \approx k_{ij} \Delta_\tau$
are observable quantities.
From the earlier discussions 
$\Delta \theta_{12} > 0$, 
so that $\theta_{12} > 35.4^\circ$ is a test for our scenario.
Another test is the compulsion of normal (inverted) mass 
ordering for  
$\theta_{23} > \theta_{23}^0$ ($\theta_{23} < \theta_{23}^0$). 
Regarding $\theta_{13}$, though the high scale value is
$\theta_{13}^0=8.9^\circ$, allowed RG evolution in our scenario
can make it anywhere between $0^\circ$ and the extant upper bound
of $13^\circ$.

The $3\sigma$ allowed range of $\theta_{12}$ is 
$\theta_{12} \in (29.3^\circ, 39.8^\circ)$ \cite{nu-fits}.
With $\Delta \theta_{12} \Delta_\tau$ necessarily positive, this
implies $0<k_{12}\Delta_\tau<4.4^\circ$.
Strong constraints then ensue on $m_0$ and $\alpha_2$,
since $\Delta \theta_{12}$ is inversely proportional to the small quantity
$\epsilon_S$. 
In Fig.~\ref{fig-mbeta}, we show the $3\sigma$
allowed values of $m_0$ and $\bb$ for two $\tan\beta$ values.
The figure is obtained by solving the RG equations numerically
with $\theta_{ij}^0$'s as the initial conditions at the high
scale, 
and marginalizing over $\alpha_3$ and $\delta$.
The figure may be understood easily
with our analytic expressions (\ref{kij-rhoeps}).
At large $\tan \beta$, the value of $\bb$
has to be close to $\pi$ in order to avoid an excessive
$k_{12}$ enhancement \cite{minakata} (even this will
not work if $m_0$ is too large).
A nontrivial Majorana phase is thus essential.
For smaller values of $\tan\beta$, however, no such tuning
is required (see fig.~\ref{fig-mbeta}).

\begin{figure}
\epsfig{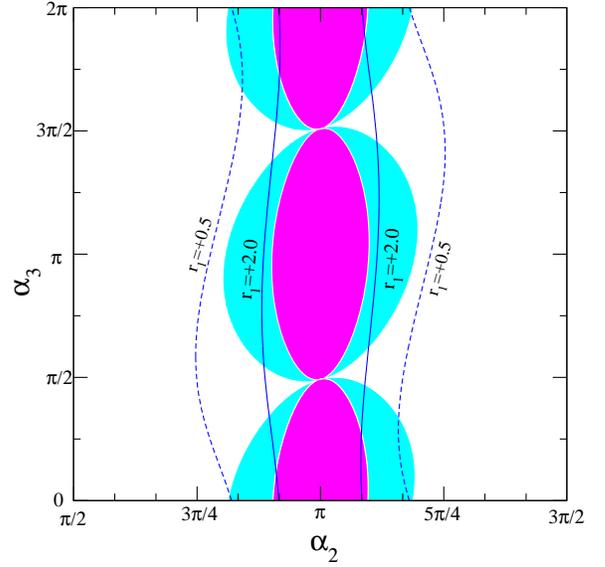}
\caption{The contours of the ratios 
$r_1 \equiv \Delta \theta_{23}/\Delta \theta_{12}$ and 
$r_2 \equiv \Delta \theta_{13}/\Delta \theta_{12}$ for
normal mass ordering. 
The line contours are for $r_1 = 2.0$ (solid) and 
$r_1 = 0.5$ (dashed). The outer edges of the cyan (light)
and magenta (dark) 
shaded regions at the centre 
correspond to $r_2 = 0.5~ (2.0)$ and 
those of the shaded regions
at the top and bottom correspond to 
$r_2 = -0.5~ (-2.0)$. 
\label{kratio}}
\end{figure}

The ratios $r_1 \equiv \Delta \theta_{23}/\Delta \theta_{12}$
and $r_2 \equiv \Delta \theta_{13}/\Delta \theta_{12}$
can be used to constrain the Majorana phases $\alpha_2$ and $\alpha_3$.
In the QDN scenario, where $\rho_A \ll 1 \ll 1/\rho_A$,
these constraints are independent of $\Delta_\tau$, and hence $\tan\beta$.
We show in Fig.~\ref{kratio} the contours of constant $r_1$ and $r_2$
in the $\alpha_2$--$\alpha_3$ plane, for normal hierarchy.
With inverted hierarchy, the signs of $r_1$ and $r_2$ are reversed. 
Note that $\alpha_2=\alpha_3=0$
necessitates $r_1 \approx 2 \delta m^2_S/\delta m^2_A \approx 0.06$ 
and $r_2=0$,
which implies $\theta_{23} \approx \theta_{23}^0 \;,
\theta_{13} = \theta_{13}^0$.
Any deviation from this prediction will indicate 
non-zero Majorana phases.
However, for these relations to be practically useful as a test,
measurements of these angles
accurate to within a couple of degrees are essential. 

If $\alpha_2 \approx \pi$, which would be the case if
$\theta_{12}$ is found to be very close to $\theta_{12}^0$,
the ratio $r_2/r_1 \approx -2 \cos \alpha_3$ gives
a direct measurement of $\alpha_3$,
with $|r_2/r_1| <2$ serving as a weak test of this scenario. 

Even without any knowledge of the Majorana phases,
the measurements of the mixing angles can put a lower bound
on $\Delta_\tau$, and hence on $\tan\beta$. 
With QDN we have the
relations
$
|\Delta\theta_{12}|  <  \Delta_\tau/(4\epsilon_S) \;,
|\Delta\theta_{23}|  < \Delta_\tau/|2\rho_A| \; ,
|\Delta\theta_{13}| <  \Delta_\tau/|4\rho_A| 
$
and additionally the combinations
$|\Delta\theta_{23} \pm \Delta\theta_{13}|
<  \Delta_\tau/|2\rho_A|$ .
Once $m_0$ is known, the best of the above lower bounds on
 $\Delta_\tau$ may be chosen to restrict $\tan\beta$ from below
via eq.~(\ref{delta-tau}).

In conclusion, a basis independent formulation of QLC 
at a high scale can be consistent with the
data even for the QDN scenario
{\it provided a nontrivial role is played by the Majorana phases}.
We have explicitly shown this numerically as well as through 
transparent analytic 
approximations for the RG evolutions of the mixing angles.
Our new results are correlated constraints on the neutrino mass scale
and the Majorana phases, as well as correlations among the 
neutrino mixing angles which can be tested by their
precise measurements. 
Specifically, one of the major predictions of our scenario
is $\theta_{12} > 35.4^\circ$.
Currently the data is consistent with this prediction
to within $1\sigma$.
A further reduction of the error in $\theta_{12}$
\cite{solar-prospects} will clarify the situation.

We thank M. Schmidt for useful discussions.
The work of S.G. is supported by the
Alexander--von--Humboldt--Foundation.
The work of A.D. is partly supported through the 
Partner Group program between the Max Planck Institute
for Physics and Tata Institute of Fundamental Research.

\end{document}